\documentclass{article}
\usepackage{spconf,amsmath,graphicx}

\usepackage{amssymb}
\usepackage{stfloats}
\usepackage{amsfonts}
\usepackage{url}
\usepackage{multirow}

\title{Graph Representation Learning for Spatial Image Steganalysis}
\name{Qiyun Liu and Hanzhou Wu
\thanks{It was supported by National Natural Science Foundation of China under Grant No. 61902235 and Shanghai ``Chen Guang'' Program under Grant No. 19CG46. Corresponding author: Hanzhou Wu (E-mail: h.wu.phd@ieee.org)}}
\address{Shanghai University, Shanghai 200444, China}

\begin{document}

\maketitle

\begin{abstract}
In this paper, we introduce a graph representation learning architecture for spatial image steganalysis, which is motivated by the assumption that steganographic modifications unavoidably distort the statistical characteristics of the hidden graph features derived from cover images. In the detailed architecture, we translate each image to a graph, where nodes represent the patches of the image and edges indicate the local relationships between the patches. Each node is associated with a feature vector determined from the corresponding patch by a shallow convolutional neural network (CNN) structure. By feeding the graph to an attention network, the discriminative features can be learned for efficient steganalysis. Experiments indicate that the reported architecture achieves a competitive performance compared to the benchmark CNN model, which has shown the potential of graph learning for steganalysis.
\end{abstract}


\section{Introduction}

Steganographic traces can be well concealed by the intrinsic noise-like components of the cover image, which are often located at the high-frequency regions. It inspires us to preferentially embed secret data into these hard-to-notice regions with an adaptive mechanism \cite{EA:paper}, or the minimal-distortion framework \cite{STC:paper}, to better resist against steganalysis. Many works are designed along this line such as \cite{HUGO:paper, HILL:paper, UNIWARD:paper}. Unlike the above-mentioned algorithms that follow the paradigm of minimizing the total embedding cost, another principle for steganographic design is to preserve the chosen model of the cover, which is provably secure with respect to the chosen  model such as \cite{ProvMB:paper}. Recent works bring closer the relationship between cost based steganography and model based steganography \cite{CostMB:paper}.

As the opposite to steganography, (image) steganalysis is to determine whether a given image contains hidden information or not. It can be divided into \emph{targeted steganalysis} and \emph{blind steganalysis}. While the former uses the prior knowledge about the specific steganographic algorithm to identify stegos created by the steganographic algorithm, the latter aims to determine whether a given media contains secret information or not without any side information about the steganographic algorithm. Targeted steganalysis usually shows better detection performance than blind steganalysis since steganographic traces are more likely to be captured given the embedding mechanism of the steganographic algorithm. However, in applications, blind steganalysis is more desirable than targeted steganalysis since it is often the case that the used steganographic algorithm is unknown to us, resulting in that the majority of steganalysis works are (semi-)blind systems.

From the perspective of system design, early steganalysis extracts manually-crafted features from media objects and then uses conventional statistical analysis tools such as support vector machine and linear discriminant analysis for classification. E.g., Markov-based features have been widely used in early image steganalysis such as \cite{Markov:paper, SPAM:paper, SRM:paper}. Though \emph{ensemble} and \emph{dimensionality reduction} can be utilized to enhance the detection performance \cite{PooledSteganalysis:bookchapter}, these algorithms rely heavily on sophisticated manual feature design and become hard to improve due to the advancement of steganography that tends to alter pixels in image regions that are hard to detect. In order to overcome this difficulty, in recent years, in-depth studies are performed on moving the success achieved by deep convolutional neural networks (deep CNNs) \cite{CNN:paper} in computer vision to image steganalysis \cite{QianNet:paper, Xu:net, Xu:ensemble, Jessica:SRNet, YeNet:paper, MTAP:paper}. These works can be briefly generalized by three phases, i.e., \emph{residual determination}, \emph{feature learning} and \emph{binary classification}. Specifically, they first filter the input image to generate residual image(s). It enlarges the signal-to-noise ratio between the noise-like stego signal and the host signal and therefore facilitates the feature learning procedure. By providing the residual image(s) to the deep CNN architecture, discriminative features can be learned and utilized for binary classification. The entire process can be implemented by an end-to-end fashion.

Recently, there is increasing interest in extending the deep learning paradigms to graph data, promoting graph neural network (GNN) to become a hot topic \cite{GNN:survey}. GNNs are essentially graph representation learning models and can be well applied to node-focused tasks and graph-focused tasks. By modelling digital images as graph data structures, many visual problems can be effectively solved with GNNs. Motivated by this point, despite the superiority of CNNs in image steganalysis, in this paper, we make the step towards GNN based image steganalysis. Our experimental results indicate that the proposed GNN based architecture achieves a competitive performance, which shows the potential of graph representation learning for image steganalysis and may inspire more advanced works.

The rest of this paper will be organized as follows. In Section 2, we detail the proposed work, followed by experimental results in Section 3. We conclude this paper in Section 4.

\begin{figure*}[!t]
\centering
\includegraphics[width=6in]{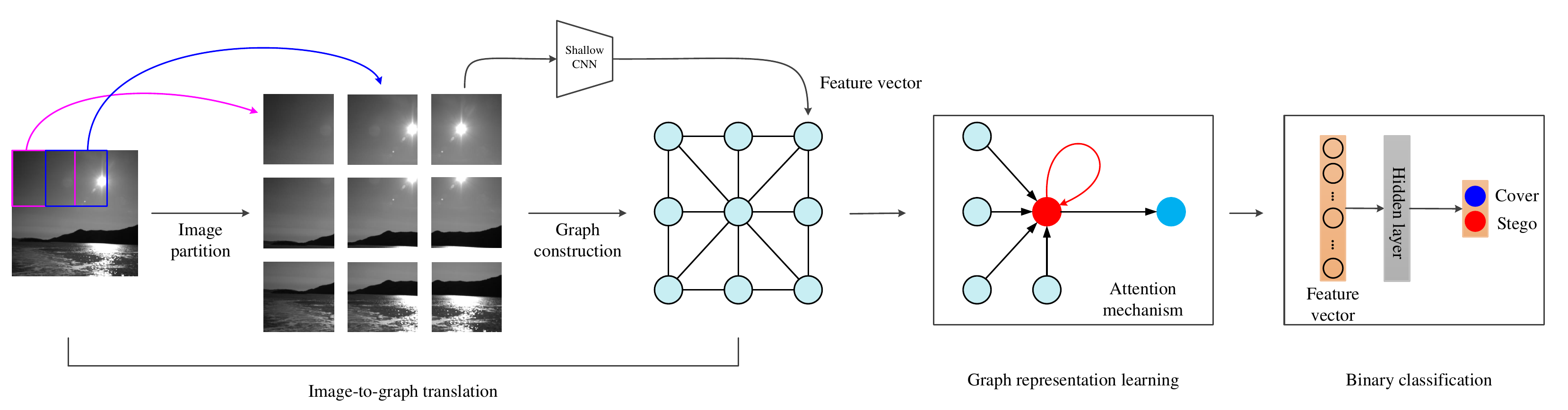}
\caption{Sketch for the proposed GNN based architecture for spatial image steganalysis.}
\end{figure*}

\section{Proposed Method}
As shown in Fig. 1, the proposed architecture consists of three phases, i.e., \emph{image-to-graph translation}, \emph{graph representation learning} and \emph{binary classification}. The purpose of image-to-graph translation is to convert an image to a graph with feature vectors assigned to the nodes. The graph can be then fed to a graph attention network for representation learning, allowing the outputted feature vector to be used for final classification.

\subsection{Image-to-Graph Translation}
Given a gray-scale image $\textbf{I} = \{x_{i,j}|1\leq i\leq h, 1\leq j\leq w\}$, where $x_{i,j}\in \{0, 1, ..., 255\}$, we first partition $\textbf{I}$ into $n\times m$ patches, where $n \leq h, m\leq w$. A patch is defined as a sub-image of $\textbf{I}$ with a size of $h_p\times w_p$, where $h_p \leq h$ and $w_p\leq w$. Let $\{\textbf{I}_{u,v}|1\leq u\leq n, 1\leq v\leq m\}$ represent the patches obtained by raster scanning, where $\textbf{I}_{u,v}$ is the patch located at position $(u,v)$, e.g., for the $3\times 3$ patches shown in Fig. 1, the central patch has a position index of $(2,2)$. The first step of image-to-graph translation is to compute all $\textbf{I}_{u,v}$ defined as:
\begin{equation*}
\textbf{I}_{u,v} = \{x_{i, j}|i\in [f_{u,v},f_{u,v}+h_p), j\in [g_{u,v},g_{u,v}+w_p)\},
\end{equation*}
where $(f_{u,v}, g_{u,v})$ represents the position of the top-left pixel of $\textbf{I}_{u,v}$ in $\textbf{I}$. Initially, we have
$f_{1,1} = g_{1,1} = 1$ and
\begin{equation}
f_{u,v} = f_{u,v-1}, g_{u,v} = g_{u-1,v}, \forall u\in [2, n], v\in [2, m].
\end{equation}
For $v\in [2,m]$, $g_{u,v}$ is determined by:
\begin{equation}
g_{u,v} = g_{u,v-1}+(1-\alpha)\cdot w_p,
\end{equation}
where $\alpha\in [0,1)$ is the parameter used to control the area of intersection between $\textbf{I}_{u,v}$ and $\textbf{I}_{u,v-1}$, e.g., $\alpha = 0.3$ means 30\% pixels in $\textbf{I}_{u,v}$ are also belonging to $\textbf{I}_{u,v-1}$. Similarly, for $2\leq u\leq n$, $f_{u,v}$ is determined by:
\begin{equation}
f_{u,v} = f_{u-1,v} + (1-\beta)\cdot h_p,
\end{equation}
where $\beta$ controls the area of intersection between $\textbf{I}_{u-1,v}$ and $\textbf{I}_{u,v}$. By default, we use $\alpha = \beta$. 

For example, assuming that $h=w=2h_p=2w_p=512$, we have 4 disjoint patches if $\alpha = 0$ and $n=m=2$. And, $(f_{1,1},g_{1,1}) = (1,1)$, $(f_{1,2},g_{1,2}) = (1,257)$, $(f_{2,1},g_{2,1}) = (257,1)$, $(f_{2,2},g_{2,2}) = (257,257)$. We have 9 patches if $\alpha = 0.5$ and $n = m = 3$. Moreover, the top-left pixel-positions are $(1,1)$, $(1,129)$, $(1,257)$, $(129,1)$, $(129,129)$, $(129,257)$, $(257,1)$, $(257,129)$ and $(257,257)$, respectively.

In order to construct a graph, each patch will be mapped to a graph node. Then, edges should be assigned to node-pairs. For example, for any two different nodes, we can add an edge between them so that a \emph{complete graph} can be generated. We can also use the spatial relationship between nodes to construct the graph. For example, for two patches $\textbf{I}_{a,b}$ and $\textbf{I}_{c,d}$, we add an edge between the corresponding two nodes if $\text{max}(|a-c|,|b-d|)=1$. We define such graph as a \emph{lattice graph}. It is always free for us to design the edge insertion mechanism, which is not the main interest of this paper.   

The graph nodes should be associated with feature vectors beneficial to steganalysis. To this end, we use a shallow CNN to reduce each high-dimensional patch to a low-dimensional feature vector that will be assigned to the corresponding node. It is also open for us to design the CNN. For simplicity, we use our previously designed XWS-CNN (co-authored with G. Xu and Y. Shi) \cite{Xu:net} for feature extraction. The XWS-CNN has a high-pass filtering layer,  five conv-pooling layers and a linear classification layer. For feature extraction, we only use the high-pass filtering layer and the conv-pooling layers, enabling each patch to be mapped to a $l$-D feature vector, where $l$ is tunable, e.g., $l=128$ for the original XWS-CNN in \cite{Xu:net}.

Though XWS-CNN itself has demonstrated superior performance in image steganalysis, we are to show in our experiments that the steganalysis performance of XWS-CNN will decline significantly by reducing the number of conv-pooling layers. However, by applying the graph representation learning strategy after reducing the conv-pooling layers, the image steganalysis performance can be well maintained, indicating that graph learning plays an important role in steganalysis. 

All patches will be processed with the same CNN. In other words, only one CNN module is trained, which has a lower computational cost and reduces the mismatch impact between multiple CNNs and multiple patches caused by diversity.

\subsection{Graph Representation Learning}
The image-to-graph translation procedure enables us to construct a graph containing $nm$ nodes, which can be expressed as two matrices $\textbf{A}\in \{0,1\}^{nm\times nm}$ and $\textbf{W}\in \mathbb{R}^{nm\times l}$. Here, $\textbf{A}$ means the adjacency matrix and $\textbf{W}$ denotes the node features in the matrix form. The purpose of graph representation learning is to use a GNN to generate a representation (embedding) for each node in the above graph so that the node embeddings in the matrix form can be used to determine whether the corresponding image is \emph{stego} or not. We have evaluated various GNNs and finally decided to use the graph attention network (GAT) \cite{GAT:paper} due to its superiority for steganalysis. Following the neighborhood aggregation paradigm, GAT takes as input a
graph (including its topological structure and descriptive features) and produces a representation for each graph node. 
The representation for each node can be expressed as a vector. We refer the reader to \cite{GAT:paper} for more details. One thing to note is that before feeding $\textbf{A}$ to the GAT, for each node, we add an edge to connect the node itself, i.e., $a_{i,i} = 1$ for all $a_{i,i}\in \textbf{A}$. In addition, \emph{multi-head attention}  \cite{GAT:paper} is not used in this paper. 

\subsection{Binary Classification}
To achieve graph-level classification, a \emph{readout} function \cite{GNN:survey} is used to generate a representation (embedding) of the entire graph according to the final node representations of GAT. In other words, the readout operation maps the final node representations in the matrix form to a real vector. The graph-level representation will be thereafter fed to a 64-D fully-connected hidden layer equipped with ReLU \cite{relu:leakyrelu}, and then processed by a 2-D classification layer with softmax to output probabilities.

\begin{figure}[!t]
\centering
\includegraphics[width=\linewidth]{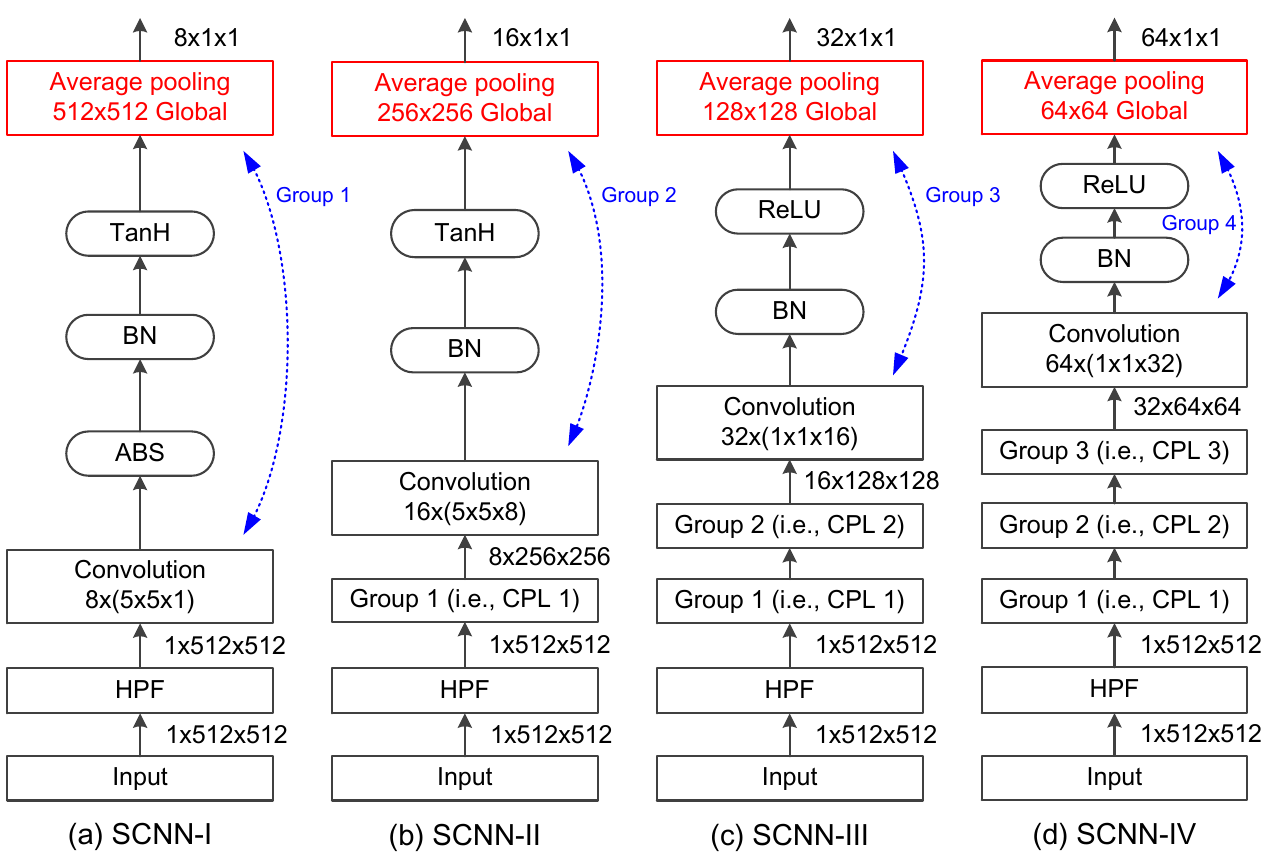}
\caption{Four different shallow CNNs based on XWS-CNN.}
\end{figure}

\section{Experimental Results and Analysis}
\subsection{Basic Setup}
The steganographic algorithms tested in experiments included S-UNIWARD \cite{UNIWARD:paper} and HILL \cite{HILL:paper}. The popular BOSSBase 1.01 \cite{HUGO:paper} dataset containing 10,000 natural images sized $512\times 512$ was used for generating stego images. For each experiment, out of 10,000 pairs of (cover/stego) images, 4,000 pairs were used for model training, 1,000 pairs were used for model validation and the rest 5,000 pairs were used for model testing. The three subsets did not intersect with each other.

We used PyTorch for simulation, accelerated with a single TITAN RTX 24 GB GPU. The batch size was 32, and the number of epochs was 300, resulting in a total of 75,000 iterations for training a model. The learning rate was 0.001 and the Adam optimizer \cite{Adam:Paper} (with two hyper-parameters $\beta_1 = 0.5, \beta_2 = 0.999$) was used to update model parameters, which were initialized with the default settings provided by PyTorch.

Our GAT implementation based on the open source code\footnote{\url{https://github.com/Diego999/pyGAT}}. In experiments, the number of graph attention layers (GALs) was 2. The readout function used the ``average'' operation to pool node representations. The dimension of the graph-level representation processed with readout was equal to that of the representation of each graph node. Both \emph{complete graph} and \emph{lattice graph} mentioned above were used for evaluation. And, $h_p = w_p = 256$, $n = m = 3$, $\alpha = \beta = 0.5$ were used.

\subsection{Shallow CNN}
We used the high-pass filtering (HPF) layer and conv-pooling layers (CPLs) of XWS-CNN \cite{Xu:net} to construct the shallow CNN. There are 5 CPLs in XWS-CNN. By orderly collecting CPLs from bottom to top, we can construct 5 different shallow CNNs. For example, a shallow CNN may only consist of the HPF layer and Group 1 (i.e., the first CPL, or say CPL 1) of XWS-CNN. To guarantee that the feature vector outputted by the shallow CNN can be fed to the subsequent layer, the \emph{last} average pooling operation was set to \emph{global}. Fig. 2 provides four shallow CNNs by applying the HPF layer and CPLs of XWS-CNN. Obviously, by applying the HPF layer and all CPLs of XWS-CNN, we can construct the fifth shallow CNN, i.e., SCNN-V. \{SCNN-I, SCNN-II, ..., SCNN-V\} can be used for steganalysis by adding the classification layer of XWS-CNN. The corresponding models are termed as SCNN-I+BC, SCNN-II+BC, SCNN-III+BC, SCNN-IV+BC, SCNN-V+BC. Here, BC is short for binary classification. Obviously, SCNN-V+BC is equivalent to XWS-CNN. We built five graph models based on \{SCNN-I, SCNN-II, ..., SCNN-V\} by applying the proposed framework. The corresponding five graph models are termed as SCNN-I+GNN+BC, SCNN-II+GNN+BC, ..., SCNN-V+GNN+BC. One thing to note is that the input size of the shallow CNN should be adjusted to $h_p\times w_p$ for the proposed method.

\begin{figure}[!t]
\centering
\includegraphics[width=3in]{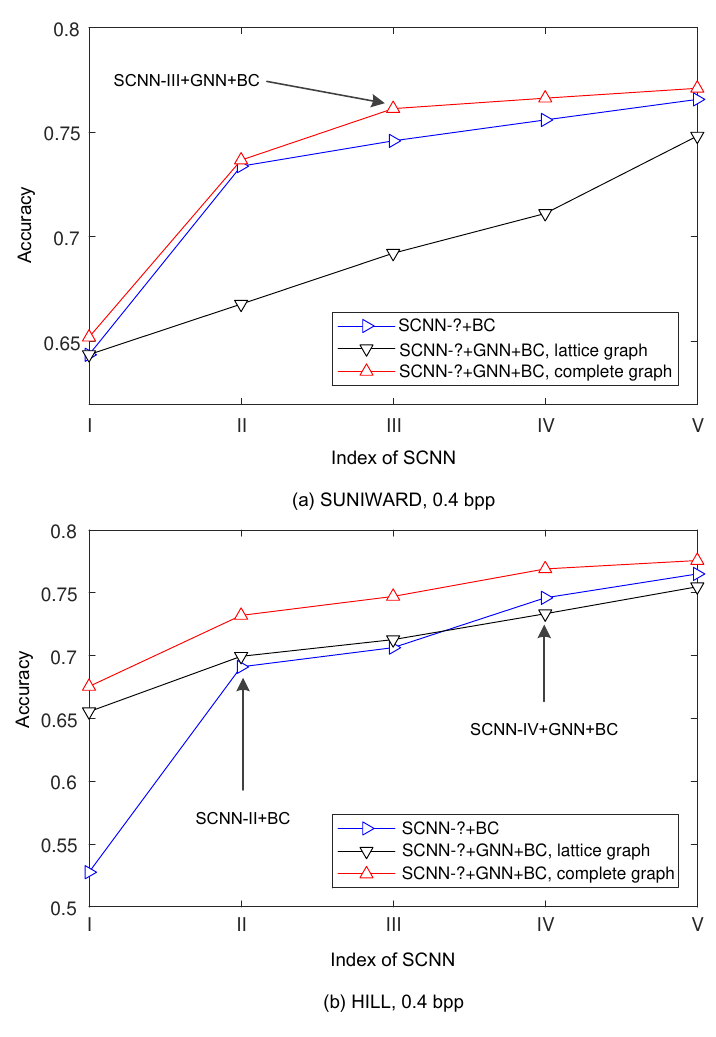}
\caption{Detection accuracy with an embedding rate of 0.4 bpp.} 
\end{figure}

\begin{figure}[!t]
\centering
\includegraphics[width=3in]{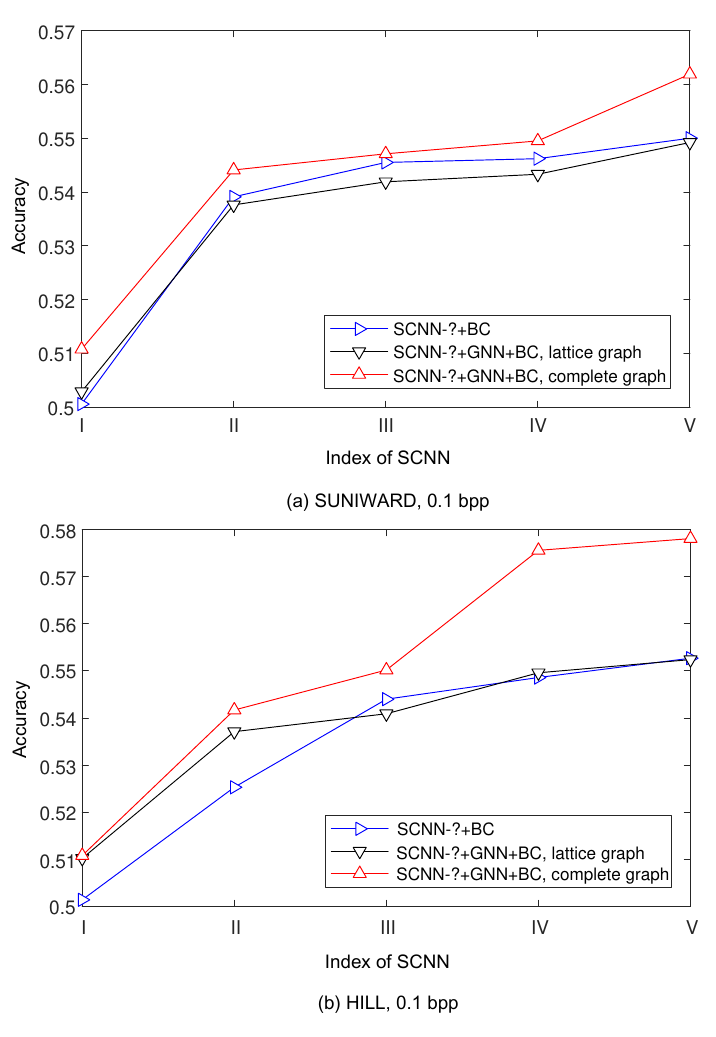}
\caption{Detection accuracy with an embedding rate of 0.1 bpp.} 
\end{figure}

\subsection{Results}
The most common indicator detection accuracy defined as the percentage of correctly classified images was used to evaluate the performance. Meanwhile, two representative embedding rates: 0.4 bpp (bits per pixel) and 0.1 bpp, were tested. Fig. 3 and Fig. 4 show the results with an embedding rate of 0.4 bpp and 0.1 bpp respectively for S-UNIWARD and HILL using different models. From the two figures, we can conclude that: First, complete graph is superior to lattice graph, which may be due to the reason that complete graph enables any two nodes to be connected to each other so that local features can be  efficiently aggregated and transformed to discriminative features for steganalysis. It also indicates that different graphs result in different performance. How to find the optimal graph structure needs further study in the future. Second, with complete graph, the proposed method significantly outperforms the benchmark CNN model in terms of detection accuracy, which shows the superiority of graph learning for steganalysis. Third, the performance degradation rate of graph learning is much lower than the benchmark CNN model. In other words, by reducing the number of CPLs, the proposed work can still achieve a relatively higher accuracy, e.g., for HILL with an embedding rate of 0.4 bpp shown in Fig. 3, the detection accuracy (based on complete graph) is 0.7321 when the number of CPLs is 2 (corresponding to SCNN-II+GNN+BC) and 0.6755 when the number of CPLs is only 1 (corresponding to SCNN-I+GNN+BC), which are significantly higher than the benchmark CNN model. It implies that graph learning has the ability to well exploit statistical features and structural information for steganalysis.

\section{Conclusion}
This paper presents a general graph learning framework for spatial steganalysis. Experiments show that it outperforms the benchmark CNN model, which has verified the superiority of graph learning. Our main contribution is to reveal that graph learning has the potential to achieve state-of-the-art detection performance. One may use ensemble, prior knowledge (e.g., probability maps of embedding), parameter tuning techniques and other tricks to further enhance the detection performance. In the future, we will apply graph learning to JPEG domain.

\end{document}